%% file: SMP-19-014_temp.tex
\pdfoutput=1

\documentclass[11pt,twoside,a4paper,cmspaper,final,collab]{cms-tdr}
\def\svnVersion{8880518-D}\def\svnDate{2020/08/20}\def\cmsCernNoTag{CERN-EP-2020-076}\def\cmsCernDate{\today}\def\cmsMessage{"Published in Physical Review Letters as \href{http://dx.doi.org/10.1103/PhysRevLett.125.151802}{\doi{10.1103/PhysRevLett.125.151802}.}"}

\begin{document}\cmsNoteHeader{SMP-19-014}

\hyphenation{had-ron-i-za-tion}
\hyphenation{cal-or-i-me-ter}
\hyphenation{de-vices}
\RCS$HeadURL$
\RCS$Id$

\newlength\cmsTabSkip\setlength\cmsTabSkip{1ex}
\newlength\cmsFigWidth
\ifthenelse{\boolean{cms@external}}{\setlength\cmsFigWidth{0.48\textwidth}}{\setlength\cmsFigWidth{0.6\textwidth}}
\ifthenelse{\boolean{cms@external}}{\providecommand{\cmsLeft}{upper\xspace}}{\providecommand{\cmsLeft}{left\xspace}}
\ifthenelse{\boolean{cms@external}}{\providecommand{\cmsRight}{lower\xspace}}{\providecommand{\cmsRight}{right\xspace}}
\providecommand{\cmsTable}[1]{\resizebox{\textwidth}{!}{#1}}
\ifthenelse{\boolean{cms@external}}{\providecommand{\CL}{C.L.\xspace}}{\providecommand{\CL}{CL\xspace}}
\ifthenelse{\boolean{cms@external}}{\providecommand{\NA}{\ensuremath{\cdots}\xspace}}{\providecommand{\NA}{\ensuremath{\text{---}}\xspace}}

\newcommand{\WWW}{\ensuremath{\PW\PW\PW}\xspace}
\newcommand{\WWZ}{\ensuremath{\PW\PW\PZ}\xspace}
\newcommand{\WZZ}{\ensuremath{\PW\PZ\PZ}\xspace}
\newcommand{\ZZZ}{\ensuremath{\PZ\PZ\PZ}\xspace}
\newcommand{\VVV}{\ensuremath{\PV\PV\PV}\xspace}
\newcommand{\MTmax}{\ensuremath{\mT^{\text{max}}}\xspace}
\newcommand{\MTthird}{\ensuremath{\mT^{\text{3rd}}}\xspace}
\newcommand{\SSee}{\ensuremath{\Pepm\Pepm}\xspace}
\newcommand{\SSem}{\ensuremath{\Pepm\PGmpm}\xspace}
\newcommand{\SSmm}{\ensuremath{\PGmpm\PGmpm}\xspace}
\newcommand{\Mll}{\ensuremath{m_{\Pell\Pell}}\xspace}
\newcommand{\ptlll}{\ensuremath{\pt^{3\Pell}}\xspace}
\newcommand{\ptlllvec}{\ensuremath{\ptvec^{\kern2pt{3\Pell}}}\xspace}
\newcommand{\ptllll}{\ensuremath{\pt^{4\Pell}}\xspace}
\newcommand{\ptllllll}{\ensuremath{\pt^{6\Pell}}\xspace}
\newcommand{\DPhilllMET}{\ensuremath{\Delta\phi\left(\ptlllvec,\ptvecmiss\right)}\xspace}
\newcommand{\Mlll}{\ensuremath{m_{3\Pell}}\xspace}
\newcommand{\Mjj}{\ensuremath{m_{\mathrm{jj}}}\xspace}
\newcommand{\MjjL}{\ensuremath{m_{\mathrm{JJ}}}\xspace}
\newcommand{\DetaJJ}{\ensuremath{\Delta\eta_{\mathrm{JJ}}}\xspace}
\newcommand{\theLumi}{\ensuremath{137\fbinv}\xspace}
\newcommand{\MZ}{\ensuremath{m_{\PZ}}\xspace}
\newcommand{\ZZ}{\ensuremath{\PZ\PZ}\xspace}
\newcommand{\WZ}{\ensuremath{\PW\PZ}\xspace}
\newcommand{\WW}{\ensuremath{\PW\PW}\xspace}
\newcommand{\VH}{\ensuremath{\PV\PH}\xspace}
\newcommand{\Wg}{\ensuremath{\PW\PGg}\xspace}
\newcommand{\ttW}{\ensuremath{\ttbar\Wpm}\xspace}
\newcommand{\tWZ}{\ensuremath{\PQt\PW\PZ}\xspace}
\newcommand{\ttZ}{\ensuremath{\ttbar\PZ}\xspace}
\newcommand{\ttV}{\ensuremath{\ttbar\PV}\xspace}
\newcommand{\ttH}{\ensuremath{\ttbar\PH}\xspace}
\newcommand{\signalmu}{\ensuremath{\mu}\xspace}
\newcommand{\signalmuWWW}{\ensuremath{\mu_{\WWW}}\xspace}
\newcommand{\signalmuWWZ}{\ensuremath{\mu_{\WWZ}}\xspace}
\newcommand{\signalmuWZZ}{\ensuremath{\mu_{\WZZ}}\xspace}
\newcommand{\signalmuZZZ}{\ensuremath{\mu_{\ZZZ}}\xspace}
\newcommand{\signalmuComb}{\ensuremath{\mu_{\text{comb}}}\xspace}
\newcommand{\SSWW}{\ensuremath{\Wpm\Wpm}\xspace}
\newcommand{\sdev}{\unit{sd}}

\cmsNoteHeader{SMP-19-014}
\title{Observation of the production of three massive gauge bosons at \texorpdfstring{$\sqrt{s}=13\TeV$}{sqrt(s)=13 TeV}}

\date{\today}
\abstract{The first observation is reported of the combined production of three massive gauge bosons ($\VVV$ with $\PV = \PW,\PZ$) in proton-proton collisions at a center-of-mass energy of 13\TeV. The analysis is based on a data sample recorded by the CMS experiment at the CERN LHC corresponding to an integrated luminosity of 137\fbinv. The searches for individual $\WWW$, $\WWZ$, $\WZZ$, and $\ZZZ$ production are performed in final states with three, four, five, and six leptons (electrons or muons), or with two same-sign leptons plus one or two jets. The observed (expected) significance of the combined $\VVV$ production signal is 5.7\,(5.9) standard deviations and the corresponding measured cross section relative to the standard model prediction is $1.02^{+0.26}_{-0.23}$. The significances of the individual $\WWW$ and $\WWZ$ production are 3.3 and 3.4 standard deviations, respectively. Measured production cross sections for the individual triboson processes are also reported.}

\hypersetup{
pdfauthor={CMS Collaboration},
pdftitle={Observation of the production of three massive bosons at sqrt(s) = 13 TeV},
pdfsubject={CMS}, 
pdfkeywords={CMS, standard model, weak interaction, gauge bosons}
}

\maketitle

The production of three massive gauge bosons \VVV (\PV = \PW, \PZ) in high energy proton-proton ($\Pp\Pp$) collisions
is interesting because the standard model (SM) predictions for these processes involve the nonabelian
character of the theory~\cite{LangackerBook}.
In particular, the presence of quadruple gauge boson interactions can be
probed through \VVV production~\cite{Belanger:1992qh,Godfrey:1995cd}.
Triple gauge boson interactions and intermediate Higgs bosons (\PH) also play a role.
If physics beyond the SM is present at mass scales not far above 1\TeV,
then cross section measurements for triple gauge boson production
might deviate from SM predictions~\cite{Buchmuller:1985,Eboli:2003nq,Degrande:2012wf,Degrande:2013kka}.
Up to now, such measurements have remained elusive because the production cross sections are low and
backgrounds are insurmountable, except for rare leptonic final states.
Next-to-leading order (NLO) SM calculations predict cross sections of
509, 354, 91.6, and 37.1\unit{fb} for \WWW, \WWZ, \WZZ, and \ZZZ production
at $13$\TeV with uncertainties of approximately
10\%~\cite{Lazopoulos:2007ix,Binoth:2008kt,Hankele:2007sb,Campanario:2008yg,Dittmaier:2017bnh}. 
These calculations include contributions from the associated production of the
Higgs boson with a \PV boson, where \PH decays to $\PWp\PWm$ or
\ZZ~\cite{Han:1991ia,Brein:2003wg,Ciccolini:2003jy,deFlorian:2016spz}.
In this analysis, these contributions generally are not the dominant ones, even though
the cross section for \VH production is relatively large, because the event
selections described below have lower acceptances for the off-shell
vector boson in $\PH\to\PV\PV^{*}$.

This Letter reports the first observation of \VVV production in $\Pp\Pp$ collisions at 13\TeV,
using a data set corresponding to an integrated luminosity of \theLumi.
Recently, the first evidence of \VVV production in  13\TeV data was reported by the ATLAS Collaboration~\cite{Aad:2019dxu}
following earlier searches for \WWW production in 8\TeV ATLAS~\cite{Aaboud:2016ftt} and 13\TeV CMS data~\cite{CMS:2019mpq}.
Five final states are considered (where $\Pell = \Pe$ or $\Pgm$):
$\PWpm\PWpm\PWmp \to \Pell^\pm\Pell^\pm2\PGn\PQq\PAQq'$,
$\PWpm\PWpm\PWmp \to \Pell^\pm\Pell^\pm\Pell^\mp 3\PGn$,
$\PWpm\PWmp\PZ \to \Pell^\pm\Pell^\mp 2\PGn\,\Pell^\pm\Pell^\mp$,
$\PWpm\PZ\PZ \to \Pell^\pm\PGn\,2(\Pell^\pm\Pell^\mp)$,
and $\ZZZ \to 3(\Pell^\pm\Pell^\mp)$.
This corresponds to five exclusive channels: two same-sign (SS) leptons with jets, three (3\Pell), four (4\Pell), five (5\Pell),
and six (6\Pell) leptons.  Searches in the dilepton and trilepton final states target \WWW production;
four-lepton events are used to search for
\WWZ production; and five- and six-lepton events are used to search for \WZZ and \ZZZ production, respectively.

The data were recorded in 2016--2018 with the CMS detector, whose central feature is a
superconducting solenoid of 6\unit{m} internal diameter providing a
magnetic field of 3.8\unit{T}. Within the solenoid volume are a silicon pixel
and strip tracker, a lead tungstate crystal electromagnetic calorimeter, and a
brass and scintillator hadron calorimeter, each composed of a barrel and two
endcap sections.  Forward calorimeters extend the pseudorapidity ($\eta$)
coverage provided by the barrel and endcap detectors.  Muons are detected in
gas-ionization chambers embedded in the steel flux-return yoke outside the
solenoid.  Events are selected using triggers~\cite{Khachatryan:2016bia} that
require two electrons, two muons, or one electron and one muon passing loose
isolation requirements and certain transverse momentum ($\pt$) thresholds.  A detailed
description of the detector and definitions of the coordinate
system are given in Ref.~\cite{Chatrchyan:2008aa}.

The CMS event reconstruction is based on the particle-flow (PF) algorithm~\cite{Sirunyan:2017ulk},
which combines information from the tracker, calorimeters, and muon systems to identify charged and neutral hadrons,
photons, electrons, and muons, known collectively as PF candidates.
Electrons and muons from \PV decays, known as prompt leptons, are selected for offline analysis
using standard criteria~\cite{Khachatryan:2015hwa,Sirunyan:2018fpa}. 
Events containing \PGt leptons decaying into charged hadrons are rejected by requiring
the absence of isolated tracks aside from selected electrons and muons. 
The PF candidates are clustered into jets using the anti-\kt algorithm with a distance parameter
of 0.4~\cite{Cacciari:2005hq,Cacciari:2008gp,Cacciari:2011ma}.
Jets with $\pt > 20\GeV$ and $\abs{\eta} < 5$  are selected for the analysis.
Defining the distance between a jet and a selected lepton by $\Delta R = \sqrt{\smash[b]{(\Delta\eta)^2+(\Delta\phi)^2}}$
where $\phi$ is the azimuthal angle, jets are rejected if $\Delta R < 0.4$.
Jets containing the decay of a \PQb~quark are identified using
the loose working point of the deep combined secondary-vertex \PQb~tagging algorithm~\cite{Sirunyan:2017ezt}.
To increase the efficiency for identifying low-\pt \PQb~hadrons not clustered into jets, a soft \PQb
tag object~\cite{Sirunyan:2017wif} is defined using a track-based secondary vertex reconstruction.

The primary $\Pp\Pp$ interaction vertex is the reconstructed vertex with the largest
summed $\pt^2$ calculated using track-based jets and the associated missing transverse momentum (\ptvecmiss), 
the negative \ptvec sum of those jets~\cite{CMSCollaboration:2015zni}. 
Track-based jets are constructed using only tracks associated with the given vertex.
In addition to the primary interaction, other $\Pp\Pp$ interactions (pileup) produce extra charged particles
and neutral energy.   Only tracks associated with the primary vertex are used.
The average neutral energy density from pileup is estimated, and subtracted from the reconstructed jet energies
and the energy sum used in calculation of lepton isolation~\cite{Cacciari:2007fd}.

The previous search for \WWW production~\cite{CMS:2019mpq} is based on sequences of requirements
called sequential cuts.  In this Letter, that approach is extended to cover all five channels.
In addition, motivated by the relatively high yields in the SS, $3\Pell$, and $4\Pell$ channels,
multivariate techniques based on boosted decision trees
(BDTs)~\cite{Breiman:1984,Narsky:2005,Roe:2004na,Khachatryan:2016ewo,chen2016xgboost}
are applied that outperform the sequential-cut analyses. Both the sequential-cut and BDT-based 
analyses are presented.

The acceptances, efficiencies, and kinematic properties of the signal and background processes
are determined using a combination of data and simulated events.
The \POWHEG 2.0~\cite{Nason:2004rx,Frixione:2007vw,Alioli:2010xd,Re:2010bp}
and the \MGvATNLO (2.2.2 and 2.4.2) generators~\cite{Alwall:2014hca} are used 
to generate \VVV signal events (including \VH), diboson (\PV\PV), and single-\PQt background events. 
The \MGvATNLO generator is used in the leading-order (LO) mode with MLM jet matching~\cite{Alwall:2007fs}
to generate SM \ttbar, \ttbar{}+X ($\mathrm{X}=\PW, \PZ, \PH$), \PW{}+jets, \PZ{}+jets, \Wg, and \SSWW events. 
The most precise cross section calculations available are used to normalize the simulated samples,
and usually correspond to either NLO or next-to-NLO accuracy~\cite{Li:2012wna,Campbell:2011bn,Czakon:2011xx,Gavin:2012sy,Campbell:2012dh,Garzelli:2012bn,Alwall:2014hca,Cascioli:2014yka,deFlorian:2016spz,Grazzini:2016swo}. 
Parton showering, hadronization, and the underlying event are modeled by \PYTHIA (8.205 and 8.230)~\cite{Sjostrand:2014zea}
with parameters set by the CUETP8M1~\cite{Khachatryan:2015pea} and CP5 tune~\cite{Sirunyan:2019dfx}. 
The NNPDF 3.0~\cite{Ball:2014uwa} and  3.1~\cite{Ball:2017nwa} parton distribution functions (PDF)
are used in the generation of all simulated samples. Pileup is simulated and
the \GEANTfour~\cite{geant4} package is used to mimic the response of the CMS detector.

The SS channel targets \WWW production~\cite{CMS:2019mpq} and
requires exactly two SS leptons with $\pt>25\GeV$ and one or more jets.
The dilepton mass $\Mll$ must exceed $20$\GeV. This channel is subdivided into
nine signal regions according to the flavors of the leptons (\SSee, \SSem, or \SSmm)
and the jet content.  Events with exactly one jet are denoted ``1J''.  Events with two or more jets are categorized
as ``\Mjj-in'' or ``\Mjj-out'' depending on whether the dijet mass for the two jets closest in $\Delta R$ is compatible
with the \PW boson mass ($65 < \Mjj < 95\GeV$).
The background processes fall broadly into three categories. The first category contains
trilepton processes with one lepton either not selected or not reconstructed (``lost'').
Such backgrounds include \WZ and \ttZ production, which typically have only one prompt neutrino in the final state;
they are reduced by requiring $\MTmax > 90\GeV$,
where $\MTmax$ is the largest transverse mass obtained from \ptvecmiss and any lepton in the event. 
The second category consists of processes with SS lepton pairs, mainly from \SSWW{}+jets and \ttW production. 
This contribution is suppressed by requiring $\MjjL<500$\GeV and $\abs{\DetaJJ}<2.5$ for the two highest-\pt jets.
The third category includes \PW{}+jets and \ttbar{}+jets production where a final-state jet or photon is misidentified
as a charged lepton, and is labeled nonprompt.  These background contributions are suppressed using strict
lepton identification and isolation requirements and by requiring $\ptmiss > 45$\GeV.
All backgrounds containing top quarks are further reduced by excluding events with \PQb-tagged jets or soft \PQb tags. 
The background due to charge mismeasurement in Drell-Yan production is relevant only for dielectron events and is
reduced to a negligible level by requiring $\abs{\Mll-\MZ} > 10$\GeV.

The 3\Pell channel, which also targets \WWW production,
is subdivided according to the number of same-flavor
opposite-sign (SFOS) lepton pairs: 0SFOS, 1SFOS, and 2SFOS.  At least one
lepton is required to have $\pt>25\GeV$, while the others must have
$\pt>20\GeV$, except in 0SFOS where all three leptons are required to have
$\pt>25$\GeV to reduce contamination from non-prompt leptons.
Events in 1SFOS and 2SFOS must contain no jets, whereas the presence
of one jet is allowed in 0SFOS.  The background sources are similar to those in
the SS category.  Events with \PQb-tagged jets are excluded to suppress
nonprompt-lepton background from processes involving top quarks.  The
contribution from triple prompt lepton backgrounds is suppressed by requiring
$\abs{\Mll-\MZ}>20\GeV$ and $\Mll>20\GeV$ for all SFOS pairs.
Additional background reduction is achieved with the following requirements: if
exactly one SFOS lepton pair is found then \MTthird, defined as the transverse mass
calculated from \ptvecmiss and the third lepton that is not one of the SFOS pair,
must be larger than 90\GeV; and, for events with no SFOS pairs,
$\MTmax > 90$\GeV  is required.  Background contributions from nonprompt leptons and
converted or misidentified photons are reduced by requiring a large \pt of the
three-lepton system $\abs{\ptlllvec}>50\GeV$, and a large azimuthal separation
$\DPhilllMET>2.5$ between \ptvecmiss and \ptlllvec.
Events with a conversion photon emitted in a \PZ boson decay are suppressed
by requiring $\abs{\Mlll-\MZ}>10\GeV$ where \Mlll is the three-lepton invariant mass.

The 4\Pell channel targets \WWZ production.  The \PZ boson is identified through its decay to
an SFOS lepton pair with $\abs{\Mll-\MZ} < 10$\GeV. 
These leptons are required to have $\pt>25\,(10)\GeV$ for the (sub)leading lepton.
The (sub)leading lepton of the remaining non-\PZ leptons must have $\pt > 25\,(10)$\GeV.
The dominant background comes from \ZZ production, so the cases of different-flavor ($\Pe\PGm$)
and same-flavor ($\Pe\Pe/\PGm\PGm$) non-\PZ lepton pairs are handled separately. 
The non-\PZ same-flavor invariant mass is required to differ from \MZ by at least 10\GeV.
Other background contributions consist of \ttZ, \tWZ, \ttH, and \WZ events.
The rejection of events with  \PQb-tagged jets reduces contributions from top quarks
and a requirement that $\Mll > 12$\GeV for all opposite-sign lepton pairs suppresses
backgrounds from low-mass resonances.  The 4\Pell channel is subdivided into seven signal regions:
for the $\Pe\PGm$ category there are four bins in \Mll and \mTii~\cite{Lester:1999tx}, and
for the $\Pe\Pe/\PGm\PGm$ category there are three bins based on \ptllll and \ptmiss. 

The 5\Pell and 6\Pell channels target \WZZ and \ZZZ production, respectively.
Event yields are low because of small cross sections and branching fractions.
Since background contributions are low, the selection maximizes the signal efficiency.
The two leading leptons are required to have $\pt > 25$\GeV and other leptons must have $\pt > 10$\GeV. 
Events in the 5\Pell channel are required to contain two SFOS lepton pairs with $\abs{\Mll-\MZ} < 15$\GeV.
The background in the 5\Pell channel consists almost entirely of \ZZ events with a nonprompt lepton,
which is usually an electron. The background is reduced by requiring $\mT>50\GeV$, where 
\mT is calculated from \ptvecmiss and that electron.
Smaller background contributions arise from \ttZ and \ttH production,
which are reduced by rejecting events with \PQb-tagged jets.
Events in the 6\Pell channel are required to have three SFOS pairs and a
six-lepton scalar \pt sum larger than 250\GeV.
The small 6\Pell background comes from \ttH and \ZZ production.

Background contributions from sources with a particular number of prompt leptons and no
nonprompt leptons in signal regions are estimated using simulations with
correction factors, typically near unity, derived from several control regions
in data enriched in the main sources of background events.  Both the predicted
numbers of events and relevant kinematic distributions are compared with
observations in control regions to derive the correction factors.  The
precision of the comparison is used to assess systematic uncertainties in these
background contributions.
Background
contributions from sources with one or more nonprompt leptons cannot be
reliably evaluated using simulations, so estimates based on control samples in data
are used instead. These estimates rely on the fact that nonprompt leptons
tend to be less isolated than prompt ones.  For the SS and 3\Pell channels,
following Ref.~\cite{CMS:2019mpq}, the
contribution of events with a nonprompt lepton is evaluated using a sample
of events in which one lepton satisfies loose identification criteria but
fails the tight criteria.  The number of events in this region determines
the estimate of the nonprompt background in the signal region using a
transfer factor computed with a separate event sample rich in nonprompt leptons.
This transfer factor is the ratio of the number of events that pass the
tight selection criteria to those that pass the loose criteria.
For the 5\Pell channel, a sample of events with three prompt leptons and one nonprompt lepton is dominated
by \WZ production and used to verify the prediction of background contributions
with nonprompt leptons.  Nonprompt leptons are a minor background for all other channels.

\begin{figure*}[htb]
   \centering
   \includegraphics[width=1.00\textwidth]{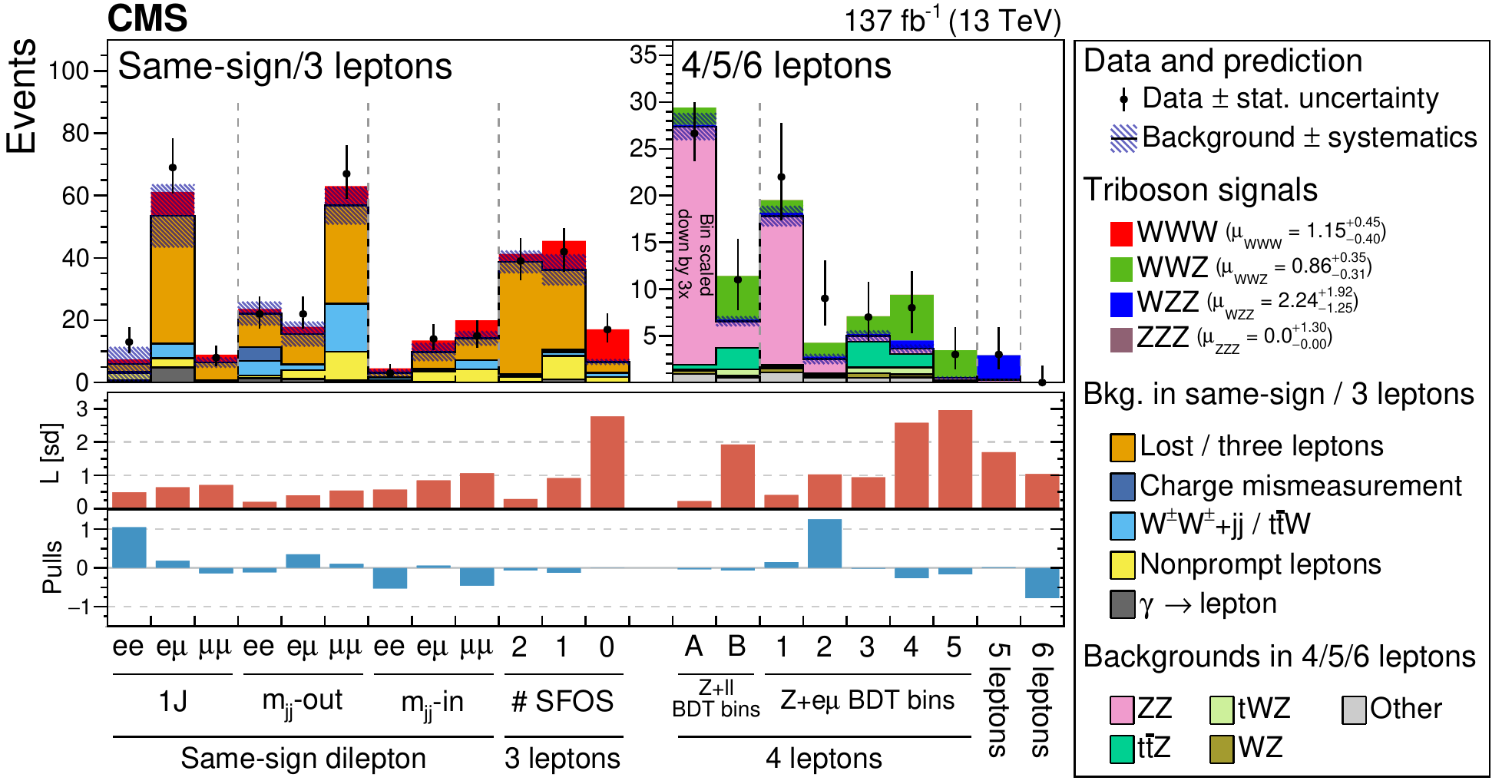}
   \caption{\label{fig:yields}
   Comparison of the observed numbers of events to the predicted yields after fitting.
   For the \WWW and \WWZ channels, the results from the BDT-based selections are used.
   The \VVV signal is shown stacked on top of the total background.
   The points represent the data and the error bars show the statistical uncertainties.
   The expected significance $L$ in the middle panel represents the number of standard deviations (sd)
   with which the null hypothesis (no signal) is rejected; it is calculated for the
   fit for $\signalmuComb$.
   The lower panel shows the pulls for the fit result.
   }
\end{figure*}

The signal strength \signalmu, defined as the measured production cross section
times branching fraction divided by the expected SM value, is determined
through simultaneous fits to all twenty-one signal regions.  In one version of
the fit, four independent signal strengths ($\signalmuWWW$, $\signalmuWWZ$,
$\signalmuWZZ$, and $\signalmuZZZ$) are used.  In the other version, a common
signal strength $\signalmuComb$ is used for all four processes.

The most important sources of systematic uncertainty involve the estimation of background contributions;
the uncertainties range from 5 to 25\% and come mainly from limited statistical precision in the
control regions.  The uncertainties in the nonprompt background estimates from
control samples in data also contribute significantly at 50\%.  Uncertainties
related to trigger efficiencies, lepton identification and energy resolution, jet
energy scale, and \PQb-jet tagging efficiency range from 1 to 9\%.  A
2.3--2.5\% uncertainty in the integrated luminosity is
assessed~\cite{REFLUMI16,REFLUMI17,REFLUMI18}.  Uncertainties due to
limitations of the theory include missing higher-order corrections (2--14\%),
PDF uncertainties (2--7\%), and the strong coupling \alpS
(1\%).  Theoretical and experimental uncertainties are correlated across
different channels.  Statistical uncertainties are much larger than systematic
ones.  The expected significance of the combined \VVV production signal based on the sequential-cut
selection is 5.4\unit{standard deviations} (sd), and the observed significance is
5.0\sdev.  The observed (expected) significances for the individual triboson
production processes are 2.5\,(2.9)\sdev for \WWW, 3.5\,(3.6)\sdev for \WWZ,
1.6\,(0.7)\sdev for \WZZ, and 0.0\,(0.9)\sdev for \ZZZ. 

\begin{figure}[htb]
    \centering
    \includegraphics[width=0.45\textwidth]{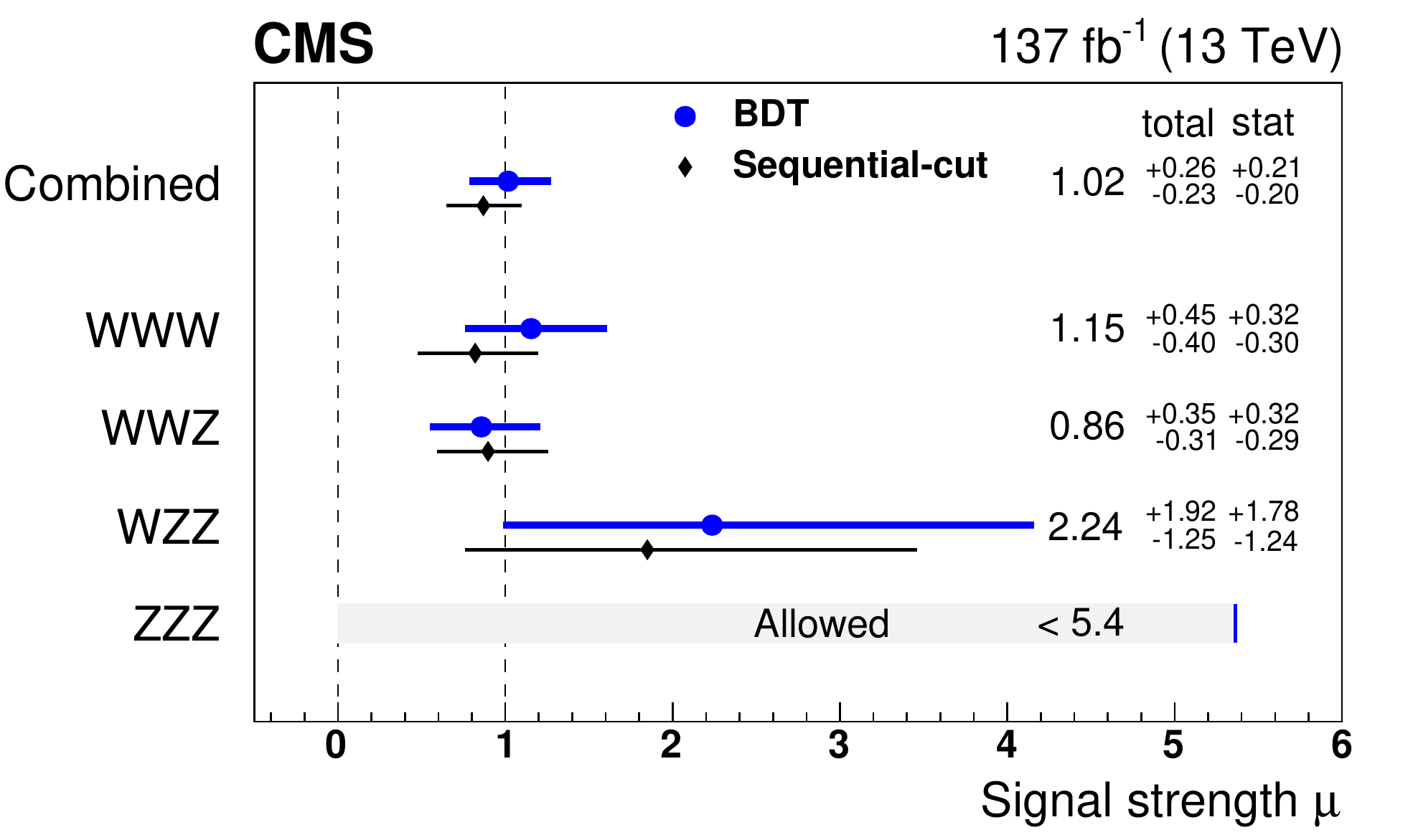}
    \caption{\label{fig:summary}
        Best fit values of the signal strengths for the BDT-based analyses (blue solid circles)
        and the sequential-cut analyses (black open circles).
        The error bars represent the total uncertainty.
        For \ZZZ production, a 95\% confidence level upper limit is shown.
        The stated numerical values correspond to the BDT-based analysis.
    }
\end{figure}

The discrimination of signal and background events in the SS, 3\Pell, and 4\Pell channels is enhanced
by using BDTs. The training and optimization of the BDTs is carried out for each channel
using simulated background and signal events.
A minimum value of each BDT output variable substitutes for the categorizations of events and
the kinematic requirements applied in the sequential-cut analyses.   In the SS and 3\Pell channels,
two separate BDTs are trained: the first one to separate signal from nonprompt background and the
second one to separate signal from the rest of the background.   These two BDTs are applied sequentially.
In the 4\Pell channel, a similar strategy is pursued except that the two BDTs are targeted against
\ZZ and \ttZ backgrounds specifically.  
There are two (five) signal regions for events in the $\Pe\Pe/\PGm\PGm$ ($\Pe\PGm$) category.  
The improvement in sensitivity due to the use of BDTs
varies channel by channel and is in the range 5--15\%.   
No BDTs are used for the 5\Pell and 6\Pell channels. 

The yields in the individual signal regions obtained using the BDTs are shown in Fig.~\ref{fig:yields}. 
The significances $L$ of the expected numbers of events are computed including systematic
uncertainties and are evaluated under the asymptotic approximation~\cite{Cowan:2010js}.
Pulls are the differences in the numbers of observed and
predicted events normalized to the uncertainties in the numbers of predicted events.
Assuming the SM production of \VVV events, the expected significance of the fit with a single
signal strength $\signalmuComb$ is 5.9\sdev and the observed significance is 5.7\sdev.
The observed (expected) significances for the individual triboson production
processes are 3.3\,(3.1)\sdev for \WWW, 3.4\,(4.1)\sdev for \WWZ, 1.7\,(0.7)\sdev for \WZZ, and 0.0\,(0.9)\sdev for \ZZZ.
In the most sensitive signal regions, approximately one third of the \VVV events come from \VH production.
The measured signal strengths, obtained in the asymptotic
approximation of the \CLs method~\cite{Cowan:2010js}, 
correspond to the total cross sections listed in Table~\ref{tab:crosssections};
leptonic branching fractions for \PW and \PZ decays come from Ref.~\cite{PDG2018}.
If \VH is considered as a background, then the combined observed (expected) significance
for $\signalmuComb$ is 2.9\,(3.5)\sdev and the measured cross sections are listed in  Table~\ref{tab:crosssections}.
For \ZZZ production, upper limits are reported at 95\% confidence level.
Signal strengths obtained using both sequential-cut and BDT-based approaches and with \VH production
counted as signal are summarized in Fig.~\ref{fig:summary}.

 \begin{table}
   \topcaption{\label{tab:crosssections}
   Measured cross sections obtained with the BDT-based analyses.
   The uncertainties listed are statistical and systematic.
   For the results listed in the upper (lower) half of the table, Higgs boson contributions
   are counted as signal (background).
   The \VVV cross section is calculated from the fit for $\signalmuComb$.
   For \ZZZ production, 95\% confidence level upper limits are reported.
   }
   \centering
   \begin{scotch}{ l r r }
     Process & Cross section (fb) \\[\cmsTabSkip]
    & \multicolumn{2}{c}{Treating Higgs boson contributions as} \\
    & Signal & Background \\
    \rule{0pt}{2.6ex}\noindent
     \VVV     & $1010^{+210}_{-200}\,^{+150}_{-120}$ & $370^{+140}_{-130}\,^{+80}_{-60}$ \\
     \WWW     &  $590^{+160}_{-150}\,^{+160}_{-130}$ & $190^{+110}_{-100}\,^{+80}_{-70}$ \\
     \WWZ     &  $300^{+120}_{-100}\,^{\phantom{1}+50}_{\phantom{1}-40}$ & $100^{\phantom{1}+80}_{\phantom{1}-70}\,^{+30}_{-30}$ \\
     \WZZ     &  $200^{+160}_{-110}\,^{\phantom{1}+70}_{\phantom{1}-20}$ & $110^{+100}_{\phantom{1}-70}\,^{+30}_{-10}$ \\
     \ZZZ     &  $<$200 & $<$80 \\    
   \end{scotch}
 \end{table}

In summary, proton-proton collision data at $\sqrt{s} = 13$\TeV recorded with the
CMS experiment and amounting to \theLumi were used to observe
the production of three massive gauge bosons.
The significance of the observation is 5.7 standard deviations (sd) with 5.9\sdev expected.
For \WWW (\WWZ) production, the observed significance is 3.3\,(3.4)\sdev compatible
with 3.1\,(4.1)\sdev expected.
Measured cross sections for \WWW, \WWZ, and \WZZ production and an upper limit
for \ZZZ production are in agreement with the expectations of the standard model.
This Letter documents the evidence for \WWW and \WWZ production and the
first observation of the combined production of three massive gauge bosons.

\begin{acknowledgments}
  We congratulate our colleagues in the CERN accelerator departments for the excellent performance of the LHC and thank the technical and administrative staffs at CERN and at other CMS institutes for their contributions to the success of the CMS effort. In addition, we gratefully acknowledge the computing centers and personnel of the Worldwide LHC Computing Grid for delivering so effectively the computing infrastructure essential to our analyses. Finally, we acknowledge the enduring support for the construction and operation of the LHC and the CMS detector provided by the following funding agencies: BMBWF and FWF (Austria); FNRS and FWO (Belgium); CNPq, CAPES, FAPERJ, FAPERGS, and FAPESP (Brazil); MES (Bulgaria); CERN; CAS, MoST, and NSFC (China); COLCIENCIAS (Colombia); MSES and CSF (Croatia); RIF (Cyprus); SENESCYT (Ecuador); MoER, ERC IUT, PUT and ERDF (Estonia); Academy of Finland, MEC, and HIP (Finland); CEA and CNRS/IN2P3 (France); BMBF, DFG, and HGF (Germany); GSRT (Greece); NKFIA (Hungary); DAE and DST (India); IPM (Iran); SFI (Ireland); INFN (Italy); MSIP and NRF (Republic of Korea); MES (Latvia); LAS (Lithuania); MOE and UM (Malaysia); BUAP, CINVESTAV, CONACYT, LNS, SEP, and UASLP-FAI (Mexico); MOS (Montenegro); MBIE (New Zealand); PAEC (Pakistan); MSHE and NSC (Poland); FCT (Portugal); JINR (Dubna); MON, RosAtom, RAS, RFBR, and NRC KI (Russia); MESTD (Serbia); SEIDI, CPAN, PCTI, and FEDER (Spain); MOSTR (Sri Lanka); Swiss Funding Agencies (Switzerland); MST (Taipei); ThEPCenter, IPST, STAR, and NSTDA (Thailand); TUBITAK and TAEK (Turkey); NASU (Ukraine); STFC (United Kingdom); DOE and NSF (USA). 
\end{acknowledgments}

\bibliography{auto_generated}

\cleardoublepage \appendix\section{The CMS Collaboration \label{app:collab}}\begin{sloppypar}\hyphenpenalty=5000\widowpenalty=500\clubpenalty=5000\input{SMP-19-014-authorlist.tex}\end{sloppypar}
\end{document}

%% file: SMP-19-014-authorlist.tex
\vskip\cmsinstskip
\textbf{Yerevan Physics Institute, Yerevan, Armenia}\\*[0pt]
A.M.~Sirunyan$^{\textrm{\dag}}$, A.~Tumasyan
\vskip\cmsinstskip
\textbf{Institut f\"{u}r Hochenergiephysik, Wien, Austria}\\*[0pt]
W.~Adam, F.~Ambrogi, T.~Bergauer, M.~Dragicevic, J.~Er\"{o}, A.~Escalante~Del~Valle, R.~Fr\"{u}hwirth\cmsAuthorMark{1}, M.~Jeitler\cmsAuthorMark{1}, N.~Krammer, L.~Lechner, D.~Liko, T.~Madlener, I.~Mikulec, F.M.~Pitters, N.~Rad, J.~Schieck\cmsAuthorMark{1}, R.~Sch\"{o}fbeck, M.~Spanring, S.~Templ, W.~Waltenberger, C.-E.~Wulz\cmsAuthorMark{1}, M.~Zarucki
% ...
\vskip\cmsinstskip
\dag: Deceased\\
1:  Also at Vienna University of Technology, Vienna, Austria\\
% ...
